\documentclass[runningheads]{svmult} 
\usepackage{makeidx}     
\usepackage{graphicx}  
\usepackage{subeqnar}  
\usepackage{multicol}  
\usepackage{physprbb}  
\makeindex    
\def\farcs{\hbox{$.\!\!^{\prime\prime}$}}
\begin{document} 
\title*{A {\tt SAURON} view of galaxies} 

\toctitle{A {\tt SAURON} view of galaxies} 
\titlerunning{A {\tt SAURON} view of galaxies} 

\author{E.K.\ Verolme\inst{1}, M.\ Cappellari\inst{1}, G.\ van de
Ven\inst{1}, P.T.\ de Zeeuw\inst{1}, \hfill\break R.\ Bacon\inst{2},
M.\ Bureau\inst{3}, Y.\ Copin\inst{4}, R.L.\ Davies\inst{5}, E.\
Emsellem\inst{2}, \hfill\break H.\ Kuntschner\inst{6}, R.\
McDermid\inst{1}, B.W.\ Miller\inst{7}, R.F.\ Peletier\inst{8} }
 
\authorrunning{E.K.\ Verolme et al.} 

\institute{Sterrewacht Leiden, Leiden, The Netherlands 
\and Centre de Recherche Astronomique de Lyon, Saint--Genis--Laval, 
     Lyon, France 
\and Department of Astronomy, Columbia University, New York, USA 
\and Institut de Physique Nucl\'eaire de Lyon, Villeurbanne, France 
\and Physics Department, University of Oxford, Oxford, UK
\and European Southern Observatory, Garching, Germany 
\and Gemini Observatory, La Serena, Chile 
\and Department of Physics and Astronomy, University of Nottingham, 
Nottingham, UK} 

\maketitle               

\begin{abstract} 
We have measured the two-dimensional kinematics and line-strength
distributions of 72 representative nearby early-type galaxies, out to
approximately one effective radius, with our panoramic integral-field
spectrograph {\tt SAURON}. The resulting maps reveal a rich variety in
kinematical structures and linestrength distributions, indicating that
early-type galaxies are more complex systems than often assumed. We
are building detailed dynamical models for these galaxies, to derive
their intrinsic shape and dynamical structure, and to determine the
mass of the supermassive central black hole. Here we focus on two
examples, the compact elliptical M32 and the E3 galaxy NGC4365. These
objects represent two extreme cases: M32 has very regular kinematics
which can be represented accurately by an axisymmetric model in which
all stars rotate around the short axis, while NGC4365 is a triaxial
galaxy with a prominent kinematically decoupled core, with an inner
core that rotates about an axis that is nearly perpendicular to the
rotation axis of the main body of the galaxy. Our dynamical models for
these objects demonstrate that two-dimensional observations are
essential for deriving the intrinsic orbital structure and dark matter
content of galaxies.
\end{abstract}

\section{The {\tt SAURON} project} 

The formation and evolution of galaxies is one of the most fundamental
research topics in astrophysics. A key question in this field is
whether early-type galaxies form very early in the history of the
universe or are gradually built up by mergers and the infall of
smaller objects.  The answer to this problem is closely tied to the
distribution of intrinsic shapes, the internal dynamics and
linestrength distributions, and the demography of supermassive central
black holes.
 
In the few past decades, it has become clear that ellipticals,
lenticulars, and spiral bulges display a variety of velocity fields
and linestrength distributions.  Two-dimensional spectroscopy of stars
and gas is essential when attempting to derive information on the
intrinsic structure. For this reason, we have built a panoramic
integral-field spectrograph, {\tt SAURON} (\cite{bac+01}), which
provides large-scale two-dimensional kinematic and linestrength maps
in a single observation.

We commissioned {\tt SAURON} on the 4.2m William Herschel Telescope on
La Palma in 1999. In low-resolution mode, the spectrograph combines a
large field-of-view (33$''\times$41$''$) with a pixel size of
0\farcs94. When the seeing conditions are good, the high-resolution
mode, with a pixel size of 0\farcs28, allows zooming in on galactic
nuclei. {\tt SAURON} observes in the spectral range of 4810--5340 \AA,
which contains the gaseous emission lines H$\beta$ and [OIII] and
[NI], as well as a number of stellar absorption features (Mg{\it b},
Fe, H$\beta$). The instrumental dispersion is $\sim$100 km/s. Between
1999 and 2002, we have used {\tt SAURON} to observe a
carefully-selected representative sample of 72 ellipticals,
lenticulars and Sa bulges, distributed over a range of magnitudes,
ellipticities, morphologies and environments (\cite{zbe+02}).

We have finalized the data reduction, have accurately separated the
emission- and absorption lines, have calibrated the line-strength
measurements, and have in hand maps of the stellar and gaseous
kinematics and linestrengths for all 48 E and S0 objects, with those
for the spirals to follow soon. The maps reveal many examples of minor
axis rotation, decoupled cores, central stellar disks,
non-axisymmetric and counter-rotating gaseous disks, and unusual
line-strength distributions (\cite{bcc+02,zbe+02}). We have also
developed new methods to spatially bin the data cubes to a given
signal-to-noise (\cite{cc03}), and to quantify the maps with Fourier
methods (\cite{cbb+01,sfz97}). This allows accurate measurements of,
e.g., the opening angle of the isovelocity contours and of the angle
between the direction of the zero-velocity contour and the minor axis
of the surface brightness distribution (\cite{cbb+01}), enabling
various statistical investigations of the entire sample of
objects.\looseness=-2

\section{Dynamical Models} 

We are constructing detailed dynamical models which fit all kinematics
and eventually even observations of the stellar line-strengths of the
galaxies in the {\tt SAURON} survey. We do this by means of
Schwarzschild's (\cite{sch79}) orbit superposition method, which was
originally developed to reproduce theoretical density distributions
(e.g., \cite{mf96,pm02,sch79,sch93,sk00}), and was subsequently
adapted to incorporate observed kinematic data in spherical and
axisymmetric geometry (\cite{czm+99,grt+03,mcz+98,rzc+97}). We have
implemented a number of further extensions including the ability to
deal with a Multi-Gaussian Expansion of the surface brightness
distribution (\cite{cvm+02,emb94,mbe92}). We have also shown that the
large data sets that are provided by instruments such as {\tt SAURON}
can be modelled without any problems (\cite{vcc+02}).

Recently, we completed the non-trivial extension to the software that
allows inclusion of kinematic measurements in triaxial geometry
(\cite{vcz03}). As in the axisymmetric case, observational effects
such as pixel binning and point-spread-function convolution are taken
into account. The chaotic orbits are dealt with in the `standard' way
(see \cite{thz01}), and the line-of-sight velocity profile is used to
constrain the models.  In the next two sections, we describe two
applications in more detail, one in axisymmetry and the other for a
triaxial intrinsic shape.

\begin{figure}[ht] 
\begin{center} 
\includegraphics[width=11.0cm]{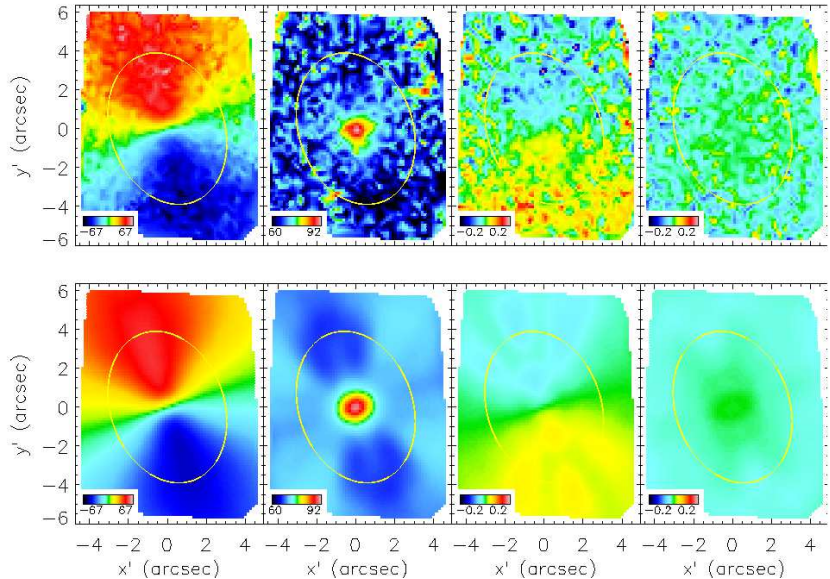} 
\end{center} 
\caption[]{Top panels: the {\tt SAURON} kinematic maps for M32. From
left-to-right: the mean velocity, velocity dispersion and
Gauss--Hermite parameters $h_3$ and $h_4$, which measure the first and
second order deviations of the line-of-sight velocity distribution
from a Gaussian shape. Bottom panels: idem, but now for the best-fit
axisymmetric dynamical model with $I$-band $M/L=1.8 M_\odot/L_\odot$,
$M_{\rm BH}=2.5 \times 10^6 M_\odot$, and $i=70^\circ$. }
\label{M32_model} 
\end{figure}

\section{Axisymmetric models for M32} 

\begin{figure}[ht] 
\begin{center} 
\includegraphics[width=12.0cm]{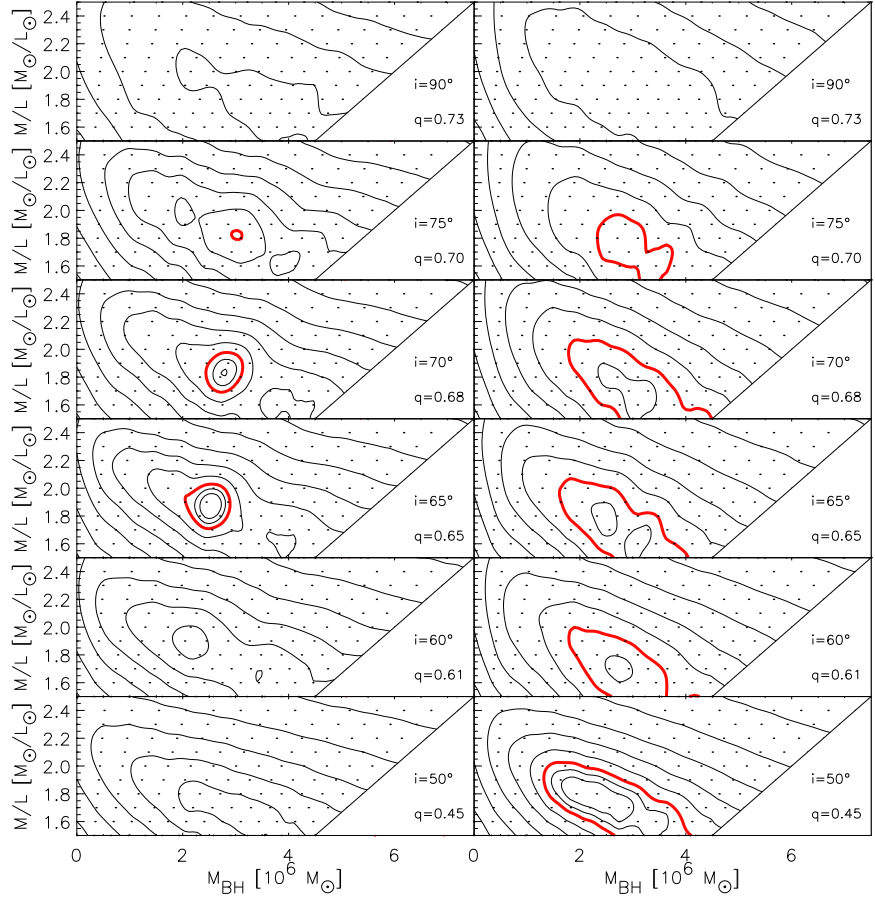} 
\end{center} 
\caption[]{Dynamical models for M32 (\cite{vcc+02}). The panels show
contours of the goodness-of-fit parameter $\Delta\chi^2$ as a function
of the central black hole mass $M_{\rm BH}$, the stellar mass-to-light
ratio $M/L$ and the inclination $i$. Each dot represents a specific
axisymmetric dynamical model.  The intrinsic flattening $q$ of the
models is indicated in the lower-right corner of each panel. The
models are constrained by {\tt STIS} kinematics along the major axis
(\cite{jmo+01}) together with two-dimensional observations obtained
with {\tt SAURON} in its high resolution mode (\cite{zbe+02}). The
inner three contours represent the formal $1,2$ and
$3\sigma$-confidence levels for a distribution with three degrees of
freedom.  {\it Left panels}: model fits to a data set consisting of
the {\tt STIS}-data and the full {\tt SAURON} field. Tight constraints
are placed on the central black hole mass and mass-to-light ratio, as
well as on the allowed range of inclinations.  {\it Right panels}: the
$\Delta\chi^2$ for models that were constrained by four extracted
slits from the $9''\times 11''$ {\tt SAURON} field (major and minor
axis, and at $\pm45^\circ$, as in \cite{mcz+98}) and the {\tt STIS}
data. This shows that the traditional kinematic coverage provides
almost no constraint on $i$, and that the resulting uncertainties on
the inferred values of $M/L$ and $M_{\rm BH}$ are correspondingly
larger. }
\label{M32} 
\end{figure}

We applied our axisymmetric modeling software to the nearby compact E3
galaxy M32 (\cite{vcc+02}). By complementing the {\tt SAURON} maps
(Figure \ref{M32_model}) with high-resolution major axis stellar
kinematics taken with {\tt STIS} (\cite{jmo+01}), the models are
constrained at both small and large radii, which allows us to measure
an accurate central black hole mass $M_{\rm BH}$, stellar
mass-to-light ratio $M/L$, {\it and} inclination $i$.  The left panels
of Figure \ref{M32} show the dependence of $\Delta\chi^2$, which is a
measure of the discrepancy between model and data, on $M_{\rm BH}$,
$M/L$ (in solar units, for the $I$-band) and $i$. The inner three
contours show the formal 1, 2 and 3$\sigma$-confidence levels for a
distribution with three degrees of freedom. The black hole mass and
mass-to-light ratio are constrained tightly at $M_\bullet=2.5\times
10^6 M_\odot$ and $M/L=1.8 M_\odot/L_\odot$, and the inclination is
constrained to a value near $70^\circ\pm5^\circ$.  The right panels of
the same figure show similar contours, but now for a data-set
consisting of the {\tt STIS}-kinematics together with four slits
extracted from the {\tt SAURON}-data.  In this case the constraints on
all three parameters, but most notably on the inclination, are much
less stringent. This demonstrates that two-dimensional observations
are essential to gain insight into the intrinsic structure of
galaxies.\looseness=-2

\clearpage

\section{The triaxial galaxy NGC 4365} 

\begin{figure}[ht] 
\begin{center} 
\includegraphics[width=\textwidth]{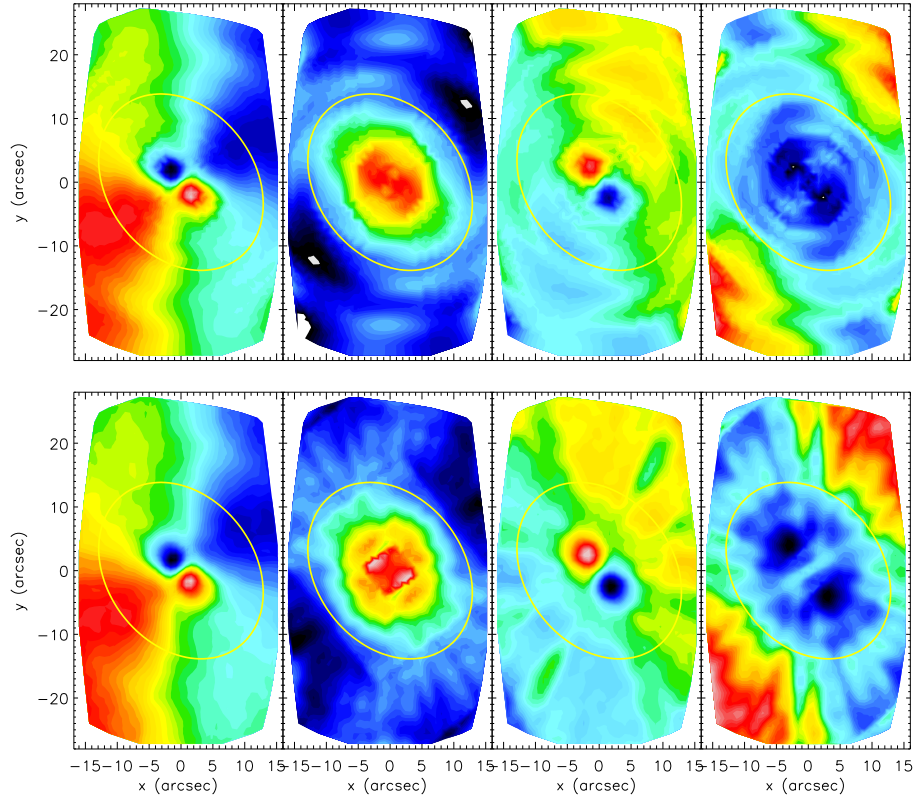} 
\end{center} 
\caption[]{Observations and dynamical models for the E3 galaxy
NGC4365. {\it Top panels}: from left to right, the stellar velocity
field, velocity dispersion, and Gauss--Hermite moments $h_3$ and
$h_4$, as observed with {\tt SAURON}. The maps are based on two
semi-overlapping pointings, sampled at $0\farcs8 \times 0\farcs8$, and
were constructed via a kinemetric expansion to provide the best
representation of the data that is consistent with an intrinsically
triaxial geometry (e.g., point-antisymmetry for the $V$ and $h_3$
maps, cf.\ \cite{cbb+01}). The original maps can be found in
(\cite{dke+01}). The amplitude of the velocity field is about 60 km/s,
the peak velocity dispersion is 275 km/s, and the contours in the
$h_3$ and $h_4$ maps range between $\pm$0.10. The decoupled core
measures $3''\times 7''$. {\it Bottom panels}: idem, but now for a
dynamical model with average intrinsic axis ratios $p=0.93$ and
$q=0.69$ (triaxiality parameter $T=(1-p^2)/(1-q^2)\sim 0.22$),
observed from a direction defined by the viewing angles
$\vartheta=85^\circ$ and $\varphi=15^\circ$. This model reproduces all
the main characteristics of the observations
(\cite{vcz+03}).\looseness=-2 }
\label{NGC4365} 
\end{figure}

The upper panels of Figure \ref{NGC4365} show the stellar kinematics
in the central $30''\times 60''$ of the giant elliptical galaxy
NGC4365, derived from two {\tt SAURON} pointings (\cite{dke+01}). The
velocity field clearly shows a prominent decoupled core in the inner
$3''\times 7''$ (cf.\ \cite{sb95}). It has a rotation axis which lies
$82^\circ\pm 2^\circ$ away from that of the body of the galaxy, which
rotates around its long axis. Such a structure is possible when the
shape is intrinsically triaxial because of the presence of orbits that
have net mean streaming around either the long or the short axis.

The globular cluster system of NGC 4365 shows evidence for an
intermediate age population (\cite{pzk+02}). The {\tt SAURON}
linestrength maps, however, indicate a predominantly old stellar
population (\cite{dke+01}), suggesting that the observed kinematic
structure may have been in place for over 12 Gyr and the galaxy is in
stable triaxial equilibrium. We therefore applied our developed
modeling software to this case, to investigate whether it is possible
to reproduce all the kinematic data in detail, and to constrain $M/L$
and the intrinsic shape and orbital structure.

We represented the observed surface brightness distribution of NGC
4365 by a Multi-Gaussian Expansion which accurately fits the observed
radial variation of ellipticity, the boxyness of the isophotes, and
the modest isophotal twisting. We derived the deprojected density by
assuming that each of the constituent Gaussian components is
stratified on similar concentric triaxial ellipsoids. The three Euler
angles that specify the orientation of the ellipsoids can be chosen
freely. For each choice, we computed a library of 4000 orbits,
obtained from 20 energy shells with 200 orbits each, covering the four
major orbit families, and including orbits from minor families and
chaotic orbits. As the spatial resolution of the {\tt SAURON}
measurements is modest, we did not consider the effect of a central
black hole. The preliminary results indicate that the quality-of-fit
parameter $\Delta\chi^2$ varies quite significantly with $M/L$ and the
parameters defining the intrinsic shape. The lower panels of Figure
\ref{NGC4365} show the predictions of one model that fits the data
well. This illustrates that the software works, and shows that NGC
4365 is indeed consistent with a triaxial equilibrium shape.

In principle, best-fit values of the shape parameters, the direction
of observation, and the mass-to-light ratio can be determined by a
systematic investigation of the parameter space, just as was done for
M32. For triaxial systems this is a very time-consuming effort, but a
first-order guess of the galaxy parameters can be obtained by using
other, simpler, schemes (see, e.g., \cite{sta94a,sta94b,vhv+03}). Our
detailed dynamical modeling software can then be used to explore this
more restricted parameter range.  Work along these lines is in
progress, and will make it possible to deduce, e.g., the intrinsic
properties of the kinematically decoupled cores seen in many of these
systems. Inclusion of higher spatial resolution data will allow
accurate measurement of the mass of the central black hole.

\clearpage
\section{Concluding remarks} 

We have presented two examples of recent results from our program to
construct detailed axisymmetric and triaxial dynamical models for
galaxies in the {\tt SAURON} representative survey of nearby
ellipticals, lenticulars and Sa bulges. The panoramic {\tt SAURON}
observations tighten the constraints on the possible orientation of a
galaxy considerably. The extension of the modeling software to
triaxial shapes including kinematic constraints works, and that it
will help us gain significant insight into the structure of early-type
galaxies.

\end{document}